# Performance-based optimal distribution of viscous dampers in structure using hysteretic energy compatible endurance time excitations

Amir Shirkhani*, Bahman Farahmand Azar**, Mohammad Ch. Basim***, Mohammadreza Mashayekhi****

**Abstract:**

*Performance-based optimization of energy dissipation devices in structures necessitates massive and repetitive dynamic analyses. In the endurance time method known as a rather fast dynamic analysis procedure, structures are subjected to intensifying dynamic excitations and their response at multiple intensity levels is estimated by a minimal number of analyses. So, this method significantly reduces computational endeavors. In this paper, the endurance time method is employed to determine the optimal placement of viscous dampers in a weak structure to achieve the desired performance at various hazard levels, simultaneously. The viscous damper is one of the energy dissipation systems which can dissipate a large amount of seismic input energy to the structure. To this end, hysteretic energy compatible endurance time excitation functions are used and the validity of the results is investigated by comparing them with the results obtained from a suite of ground motions. To optimize the placement of the dampers, the genetic algorithm is used. The damping coefficients of the dampers are considered as design variables in the optimization procedure and determined in such a way that the sum of them has a minimum value. The behavior of the weak structure before and after rehabilitation is also investigated using endurance time and nonlinear time history analysis procedures in different hazard levels.*

## 1. Introduction

Energy dissipation systems diminish the structural responses, and consequently, a better performance level is attained. Among the various types of energy dissipation devices, viscous dampers (VDs) are the most widely used devices in earthquake engineering. Such dampers also provide many advantages for seismic retrofitting of existing structures [1]. Takewak [2] investigated the effect of different arrangements of the dampers on the performance of the structure for the first time. In recent years, many types of research have been conducted on the optimal distribution of damping coefficients for viscous dampers in buildings [1, 3, 4].

\* PhD, Department of Structural Engineering, Faculty of Civil Engineering, University of Tabriz, Tabriz, Iran.
\*\* Corresponding Author: Associate professor, Department of Structural Engineering, Faculty of Civil Engineering, University of Tabriz, Tabriz, Iran. Email: b-farahmand@tabrizu.ac.ir
\*\*\* Assistant Professor, Faculty of Civil Engineering, Sahand University of Technology, Tabriz, Iran.
\*\*\*\* Assistant Professor, Faculty of Civil Engineering, K. N. Toosi University of Technology, Tehran, Iran.

The endurance time (ET) method is a dynamic analysis procedure in which structures experience intensifying dynamic excitations. This method predicts structural responses with respect to the relationship between intensity measures (IMs) and engineering demand parameters (EDPs). The EDP represents the structural responses, while IMs are related to the ground shaking intensity at various seismic hazard levels. In the ET method, a single nonlinear time history analysis (THA) offers the structural performance for a continuous range of IMs, while results of the conventional THA are just valid for a single level of IM. Indeed, structural responses at different IMs levels, presented by an incremental dynamic analysis (IDA), are provided by the ET method utilizing the least number (commonly three) of analyses [5]. In recent years, many studies have been carried out on the ET method [6-11]. Estekanchi and Basim [3] used an endurance time excitation function (ETEF) from the second generation of endurance time records [12, 13] to find the optimal distribution of VDs in structures to satisfy two hazard levels of ASCE/SEI 41-06 [14]. They investigated the performance of the structures after rehabilitation by optimally distributed VDs. Because the damage induced on a structure is a function of absorbed hysteretic energy and maximum displacement, Mashayekhi et al. [15] considered hysteretic energy compatibility in the generation process of the fifth generation of ETEFs called "ETA20kd set".

In this paper, to achieve the correct values of energy dissipated by VDs, "ETA20kd set" is used for seismic assessment of a weak steel frame rehabilitated by optimally distributed dampers. For this purpose, the genetic algorithm (GA) is utilized as the optimization tool along with the endurance time analysis (ETA). A steel frame with weaknesses in the initial design and nonlinear behavior is modeled using OpenSees software [16]. Using GA, the distribution of dampers along the height of the structure is allocated in such a way that the structure satisfies allowable limits of the code in two performance levels of LS and CP, with the minimum sum of requisite damping coefficients of the dampers. Besides, the performance of the structure in these levels is evaluated based on the percentage of dissipated energy, and interstory drift ratio (IDR) before and after the installation of VDs employing a suite of ground motions.

## 2. Endurance Time (ET) method

The concept of the ET method can be described using a hypothetical shaking-table test. Comparing the relative performance of three structures subjected to ground shaking and determining the seismic endurance is the purpose of this hypothetical test. The structures are placed on a shaking-table and subjected to an intensifying artificial dynamic excitation [5], as shown in Fig. 1.

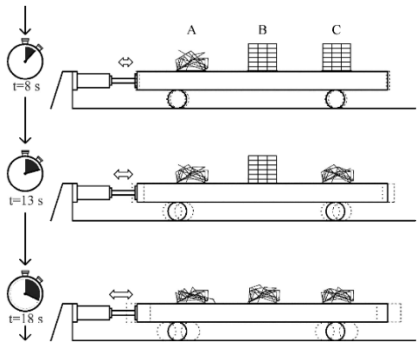

**Fig. 1:** Hypothetical shaking-table test [5].

As the amplitude of the ET excitation rises with time, the structures are expected to reach gradually from an elastic state to a nonlinear damage state and ultimately collapse. Damage indexes such as the maximum IDR of these structures are monitored during the hypothetical test of the shaking-table depicted in Fig. 1, and the results are displayed through an endurance time curve. Significant diminution of the expected computational demand is the foremost advantage of the ET method over the conventional THA procedure employing ground motions [17]. It is meriting to note that this method estimates the actual EDPs and the final design should commonly be verified utilizing more precise procedures such as IDA or cloud analysis [18, 19]. Fig. 2 shows the general methodology for utilizing the ET method [20].

## 3. Selected ground motions and ETEFs

An important factor in the success of the ET method is the availability of appropriate ETEFs. Mashayekhi et al. [15] developed a distinct simulation method of ET excitations in which hysteretic energy compatibility is considered. The previous methodologies for simulation of ETEFs include just frequency content and amplitude of motions. The duration-related parameter like ground motion duration and the factor standing for the hysteretic energy of the system can have a considerable impact on the nonlinear dynamic response of the structures [21-24]. The consistency of hysteretic energy as a parameter related to cumulative damage is considered in the process. Mashayekhi et al. [15] used the far-field earthquake records offered by FEMA P695 [25] in the simulation of the fifth generation of ETEFs called "ETA20kd set". Individual earthquake records are normalized by their corresponding peak ground velocities (PGVs). The procedure adopted in FEMA P695 has been used in the simulation of these ETEFs. Normalization by PGV is a simplistic way to eliminate unwarranted variability among earthquake records owing to the inherent differences in source type, site condition, and event magnitude while preserving inherent stochastic variability for estimating the structural response. The first horizontal components of the mentioned far-field earthquake records have been used in the simulation of the fifth generation of ETEFs. In this procedure $S_a(t,T)$ is the acceleration spectra of ETEF at time $t$ and in period $T$, which is determined as follows [15]:

$$S_a(t,T) = max\left(|\ddot{u}(\tau) + a_g(\tau)|\right) ; \quad 0 \leq \tau \leq t \quad (1)$$

where $\ddot{u}(\tau)$ is the relative acceleration response of single degree of freedom (SDOF) system with a period of T and damping ratio of 5% under ETEF, and $a_g(\tau)$ is the acceleration time history of ETEF. Then, $S_{aC}(t,T)$ indicates target acceleration spectra of ETEFs at time $t$ and in period $T$, which is obtained as follows [15]:

$$S_{aC}(t,T) = {t}/{t_{T\arg et}} \times S_a^{T\arg et}(T) \quad (2)$$

where $S_a^{T\arg et}(T)$ is target acceleration spectrum, i.e., the average response spectrum of normalized ground motions.

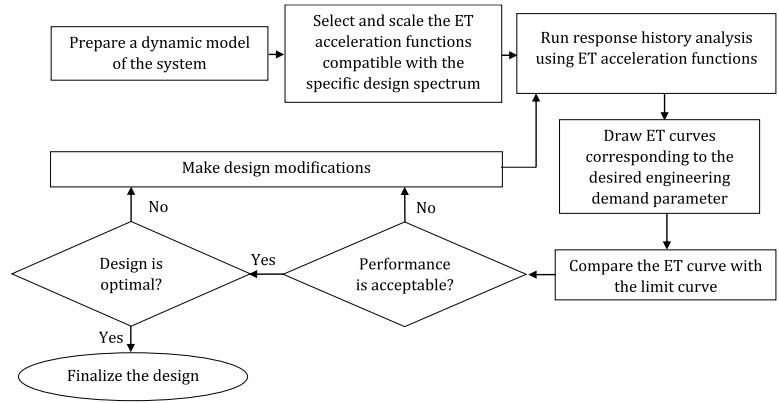

**Fig. 2:** General methodology for utilizing the ET method in optimal design procedure [20].

$t_{T\arg et}$ is the time at which the response spectra of ETEFs must match the average response spectrum of normalized ground motions. Fig. 3 illustrates $S_a^{T\arg et}(T)$ associated with the first horizontal components of 22 far-field earthquake records provided in FEMA P695. Acceptable compatibility between the response spectrum of ETA20kd01and target acceleration spectrum is shown in Fig. 4, where the response spectrum of ETA20kd01is compared to the target acceleration spectrum at $t = 5\,\text{sec}$, $10\,\text{sec}$, and $15\,\text{sec}$ in turn.

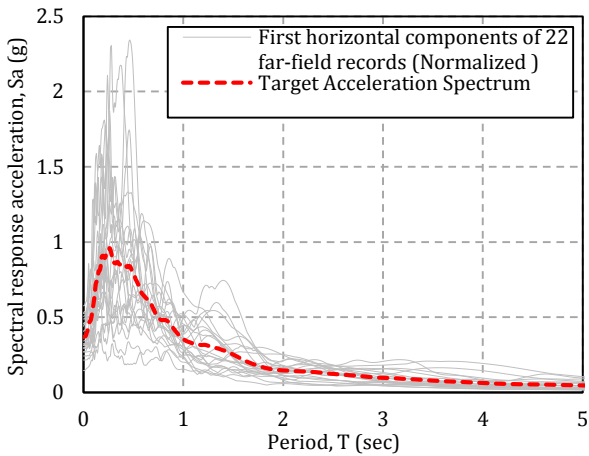

**Fig. 3:** Target acceleration spectrum, $S_a^{T\arg et}(T)$.

The first excitation function, ETA20kd01, is shown in Fig. 5. In this research, the steel frame is analyzed as a planar structure subjected to a single horizontal component of the earthquake records. Therefore, ground motions are scaled individually instead of being scaled as pairs [26]. To examine the sensitivity of the ET results to the selected earthquake records, another way to select the horizontal components of ground motions, i.e., considering maximum peak ground acceleration (PGA), is used in this study. The response spectra of these components of ground motions are exhibited in Fig 6.

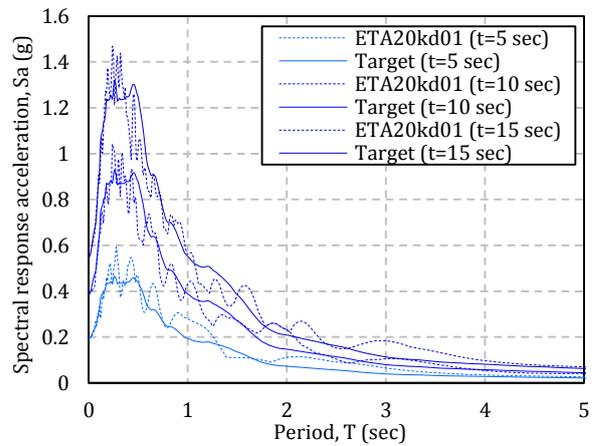

**Fig. 4:** Comparison between average response spectrum of ETA20kd01-05 and target acceleration spectrum at 5, 10, and 15 sec.

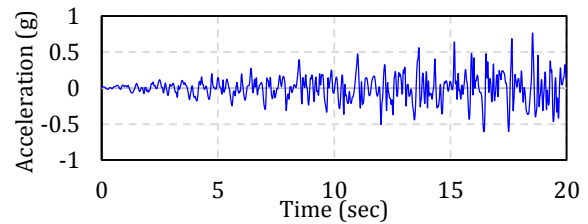

**Fig. 5:** ETA20kd01 excitation function.

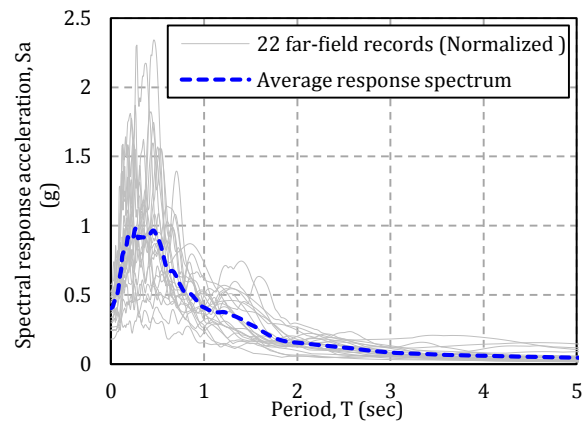

**Fig. 6:** Response spectra of horizontal components (with maximum PGA) of ground motions in this study.

## 4. Structural model and viscous dampers (VDs)

In this research, a weak three-story single-bay steel frame is considered. This structure without VDs does not satisfy the expected levels of performance of ASCE/SEI 41-06 [14] under some of the selected records. This frame has been designed assuming one-half of the base shear recommended by Standard No. 2800-05 [27] for a high seismicity area. The total mass of the frame, mass participation of first mode, the period of free vibration, and design base shear are 109.48 ton, 84.55%, 1.14 sec, and 73.9 kN, respectively [26]. This structure is selected from the study by Estekanchi et al. [26]. The geometry of this model and section properties are depicted in Fig. 7.

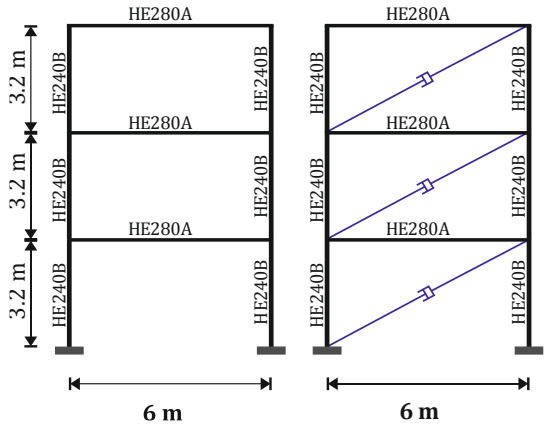

**Fig. 7:** Schematic of weak steel frame under investigation: (a) without VDs; (b) with VDs.

The height of each story is 3.2 m and the length of each bay is 6 m. Loading is performed per the Iranian National Building Code (INBC) section 6. The steel material's behavior is regarded elastoplastic with the elastic modulus of $E = 206\ GPa$ and yielding stress of $Fy = 235.44\ MPa$. The post-yield stiffness is considered 3% of the primary elastic stiffness. OpenSees software [16] is employed to perform nonlinear dynamic analyses. Beams and columns are modeled using 'nonlinearBeamColumn' element with distributed plasticity. The VDs are widely utilized in mechanical systems. The force induced in VDs depends on velocity. In this research, viscous material available in OpenSees is employed to model VDs as bracing. The damper force ($F_D$) is obtained as follows [28]:

$$F_D = sgn(\dot{x}_D) c_\alpha |\dot{x}_D|^\alpha \quad (3)$$

where $\dot{x}_D$ is the damper velocity, $c_\alpha$ is the damper coefficient and $\alpha$ is the power factor.

Soong and Constantinou [29] showed that the work done (energy dissipated by VD) in a single cycle of sinusoidal loading can be calculated as follows [28]:

$$W_D = \int_0^{T_0} F_D \dot{x}_D dt \quad (4)$$

where $T_0 = 2\pi/\omega_0$, $\dot{x}_D = x_0 \sin \omega_0 t$, and $\omega_0$ is natural circular frequencies.

## 5. Analysis and optimization procedure

Heuristic methods like GA and evolutionary algorithms are robust tools in dealing with specific structural optimization problems. The ability of these procedures in practice to obtain the optimized design utilizing codes is much more than conventional techniques because complexities like geometric properties and nonlinear dynamic characteristics do not impose any limitation in the applicability of these procedures, each type of loading, and parameters. Various constraints of design can be incorporated in the optimization method without any requirement to change the optimization algorithm. In the present study, GA which is a heuristic procedure has been utilized [30]. In this procedure, since the structural model, loadings and analyzer are independent of the optimization algorithm, required complexities in the modeling of structures and loadings can be regarded. The damping coefficient of needed VDs for the different stories of a structure is considered as the design parameter. The design criteria of the steel structure of ASCE/SEI 41-06 is used as design constraints. The minimal damping coefficient for the dampers is allocated using GA to reach the allowable limits of ASCE/SEI 41-06. Because GA does not require particular derivatives and initial assumptions and is also a random process, it can search the global solution space with more probability than other classical methods to find the general answer and does not require the objective function to possess a compatible behavior. Due to the characteristics of the problem, the behavior of the structure under dynamic loading does not follow a particular trend with respect to the design parameter, i.e., damping effect in the stories. It is worth noting that increasing damping does not necessarily result in a reduction of ground shaking effects and displacements of the structure. Hence, it seems that random search optimization is a reasonable selection in this problem.

## 6. Analysis and optimization results

### 6.1. Scaling the ground motions and the ETEFs

The period of free vibration for the weak three-story structure is 1.14 sec. Matching the response spectra at the fundamental period ($T$) of structure or considering period range ($0.2T$ to $1.5T$) can be conventionally adopted for scaling records to the desired earthquake hazard levels [31]. In this research, the spectral value at the fundamental period ($T$) of the structure is regarded. To this end, two earthquake hazard levels of ASCE/SEI 41-06 (BSE-1 and BSE-2) are considered. The equivalent ET time of ETEFs ($t_{ET}$) and the corresponding scale factors for the average response spectrum of 22 far-field earthquake records used in this study for the studied structure are tabulated in Table 1.

**Table 1.** Scale factors based on average response spectrum of ground motions and equivalent ET times for studied structure.

| Hazard level | ET time (sec) | Corresponding scale factor |
|---|---|---|
| BSE-1 | 15.4 | 1.31 |
| BSE-2 | 19.9 | 1.82 |

*6.2. ET analysis (ETA) and optimization*

In this research, according to ASCE/SEI 41-06, the design objectives 'p' and 'k' are allocated as the rehabilitation objectives. Accordingly, the structure should satisfy the Life Safety (LS) performance level in BSE-1 hazard level, and Collapse Prevention (CP) performance level in BSE-2 hazard level. These objectives are known as "basic safety objective (BSO)". According to ASCE/SEI 41-06, the allowable transient IDR is 2.5% for the LS level, and 5% for the CP level. By drawing the performance curve (ET curve) and comparing with ASCE/SEI 41-06 allowable limits, the vulnerability of each structure can be detected [32]. In this step, utilizing GA with the aim of achieving the acceptable performance expressed in ASCE/SEI 41-06, which is equivalent to being the ET curve below ASCE/SEI 41-06 allowable limit presented in Fig. 9, the requisite damping for the dampers in the structure is determined (Fig. 8). Fig. 8 (a) illustrates the total damper coefficient for the structure in each call of GA. As shown in Fig. 8 (b), the requisite damping for installed VD in the 2nd story of the studied structure is greater than the other stories. Installed VD in this story has the greatest influence on the behavior of the structure.

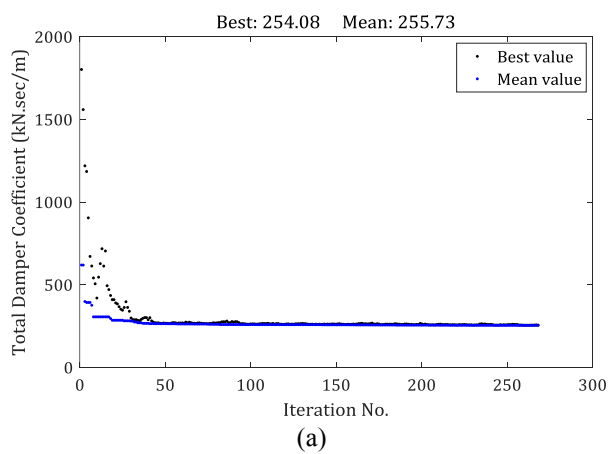

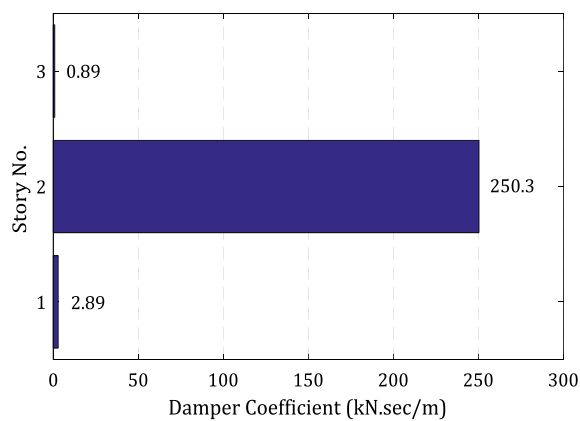

**Fig. 8:** (a) Total damper coefficient in sequential steps of GA; (b) Optimal distribution of VDs for studied structure.

The damping coefficients of the dampers are determined in a continuous range so that the performance of the studied structure after rehabilitation can be perfectly investigated, and the capability of ET method on the assessment of the behavior of the structure can be observed. In practice, discrete values can be chosen based on available commercially dampers, and VDs can be omitted from stories requiring negligible damping. The performance curves of the studied structures before and after rehabilitation are shown in Fig. 9. As illustrated in the figure, the performance curve of the structure after rehabilitation (with optimally distributed VDs) satisfies ASCE/SEI 41-06 allowable limit for LS performance levels corresponding to hazard level of BSE-1, respectively. It should be noted that this optimization is considered for "BSO". Therefore, the performance curve of rehabilitated structure is much lower than the performance curve of the weak structure.

*6.3. Validation of ET results and comparative study*

To investigate the behavior of the structure based on ground motions and to compare them with the ET results, the average of maximum IDRs of the studied structure for THA and ETA before and after the rehabilitation is compared in Fig. 10. As exhibited in this figure, the ET estimation of maximum IDRs compared to another set of the horizontal components of the ground motions is also satisfactory. Therefore, the sensitivity of maximum IDRs to another set of earthquake records is relatively low. It is also observed that after optimal distribution of VDs in the stories of structure, its performance improves. Total energy dissipated by VDs in studied structure, and the hysteretic response of VD in the 2nd story under ETA20kd01 and KOBE/SHI000 record are shown in Figs. 11 and 12, respectively. As seen in these figures, VDs dissipate a large amount of earthquake input energy. It is worth noting that time in ETA corresponds to intensity level and is somewhat different from the time in conventional THA. To prevent confusion of respected readers, it is essential to add this explanation that Fig. 12 is presented to show the performance of VDs, and not to compare the results obtained based on the ETEF and real ground motion. Tables 2-4 presents the difference between the results obtained by ETA and THA. As it is obvious from Tables 2-3, the difference between the maximum IDRs obtained by ETA and THA is greater in BSE-2 hazard levels. Undoubtedly, a more accurate estimate of these differences requires further research by increasing the number of stories and using different sets of ground motions. To have a more precise comparison between the energy values obtained by ETA and THA, the percentage of total energy dissipated by VDs for two hazard levels (BSE-1 and BSE-2) is presented in Fig. 13. These differences for the ETA results compared to THA results are equal to (-11.52%) and (-0.91%) for the earthquake hazard levels with return periods of 475 and 2475 years, respectively. In this way, the differences seem more reasonable.

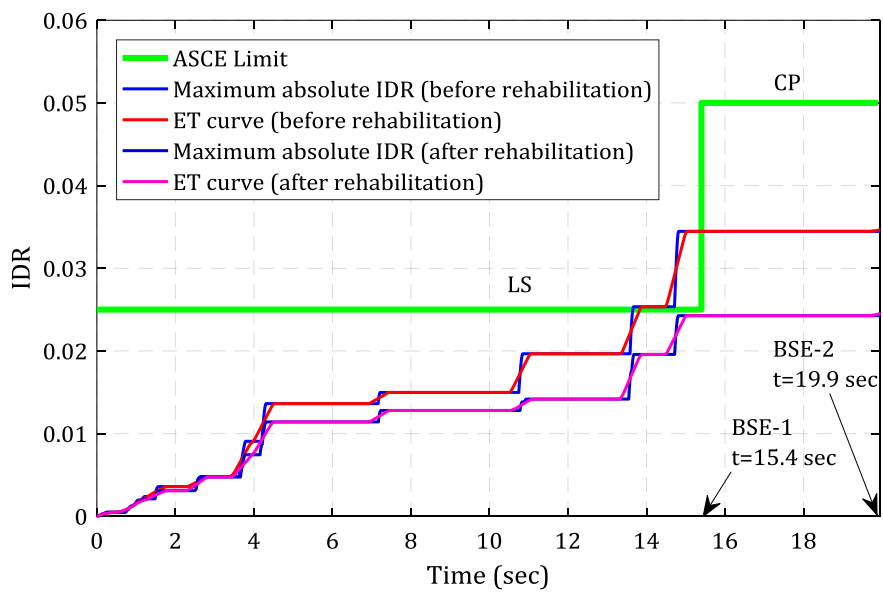

**Fig. 9:** Performance curve for studied structure before and after rehabilitation in two hazard levels.

**Table 2.** Difference between maximum IDR obtained from average response of weak structure to ETEFs and ground motions.

| Analysis | BSE-1 | BSE-2 |
| --- | --- | --- |
| THA (Ave) | 0.0229 | 0.0350 |
| ETA (Ave) | 0.0251 | 0.0284 |
| Difference (%) | 9.95 | -19.01 |

**Table 3.** Difference between maximum IDR obtained from average response of rehabilitated structure to ETEFs and ground motions.

| Analysis | BSE-1 | BSE-2 |
| --- | --- | --- |
| THA (Ave) | 0.0199 | 0.0317 |
| ETA (Ave) | 0.0208 | 0.0241 |
| Difference (%) | 4.61 | -24.07 |

**Table 4.** Difference between energy values obtained from average response of rehabilitated structure to ETEFs and ground motions.

| Energy (kN.m) | Analysis | BSE-1 | BSE-2 |
| --- | --- | --- | --- |
| Input Energy | THA (Ave) | 111.57 | 201.63 |
| | ETA (Ave) | 113.91 | 181.22 |
| | Difference (%) | 2.09 | -10.12 |
| Total Dissipated Energy (by VDs) | THA (Ave) | 44.50 | 69.02 |
| | ETA (Ave) | 40.19 | 61.47 |
| | Difference (%) | -9.67 | -10.94 |

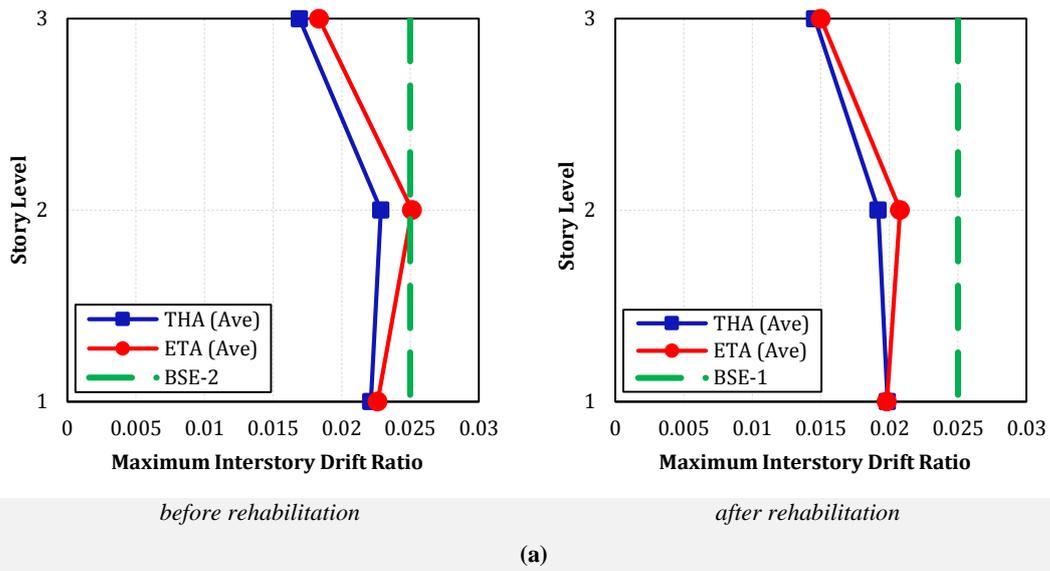

*before rehabilitation*      *after rehabilitation*

**(a)**

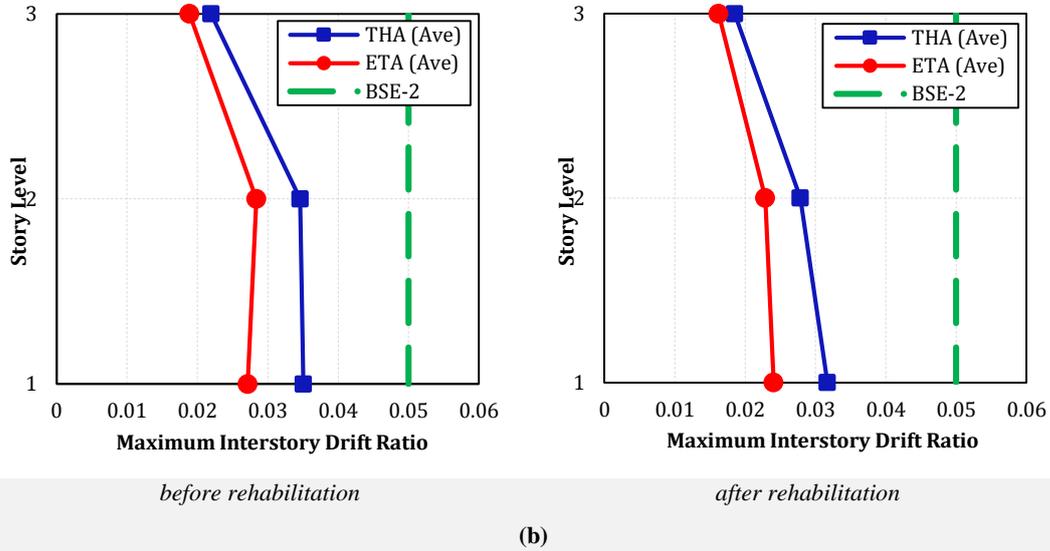

*before rehabilitation*      *after rehabilitation*

**(b)**

**Fig. 10:** Average of maximum IDRs for studied three-story frame based on THA and ETA before and after rehabilitation in: (a) BSE-1 and (b) BSE-2 hazard levels.

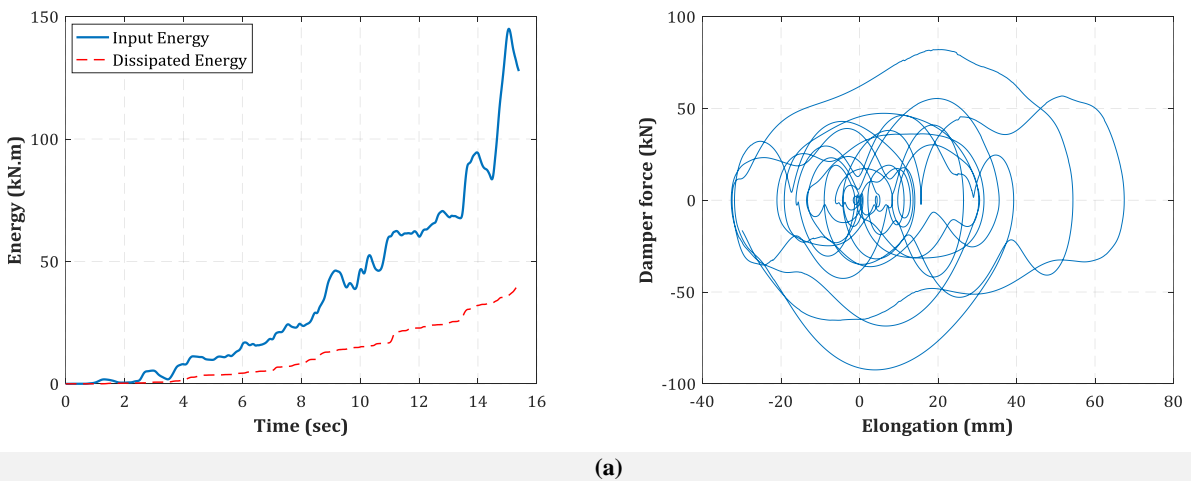

**(a)**

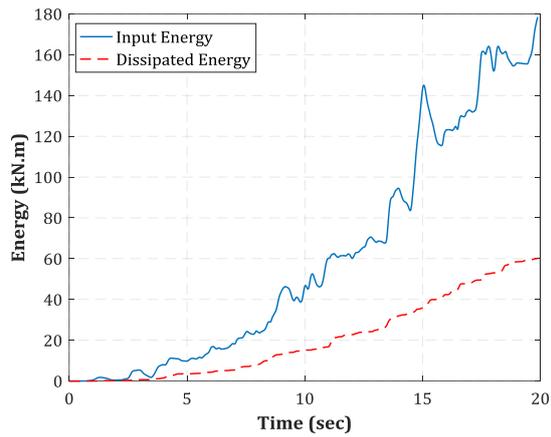
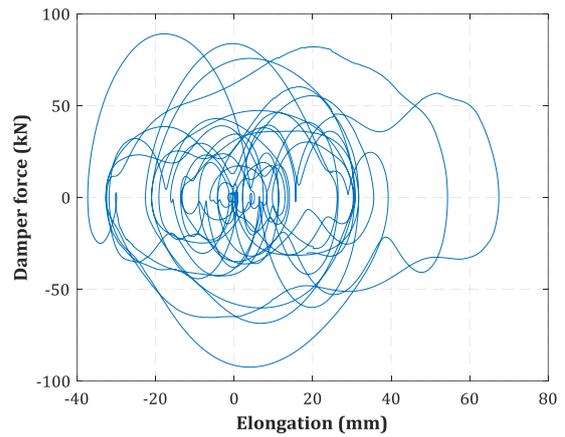

**(b)**

**Fig. 11:** Total energy dissipated by VDs in studied structure, and hysteretic response of VD in the 2nd story under ETA20kd01: (a) BSE-1 and (b) BSE-2 hazard levels.

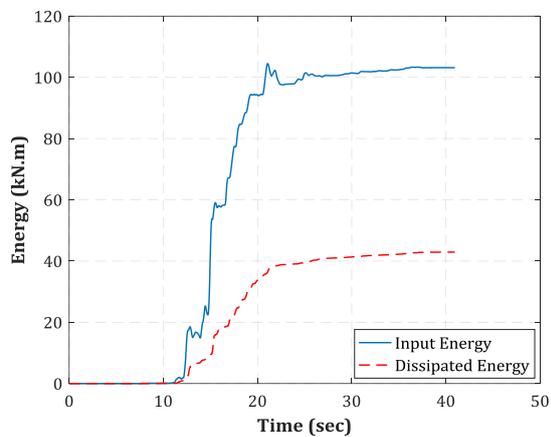
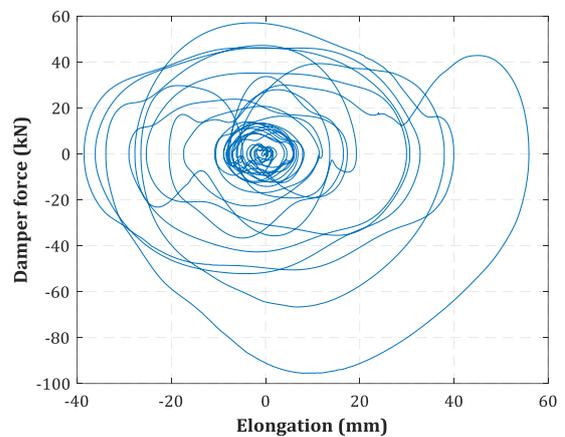

**(a)**

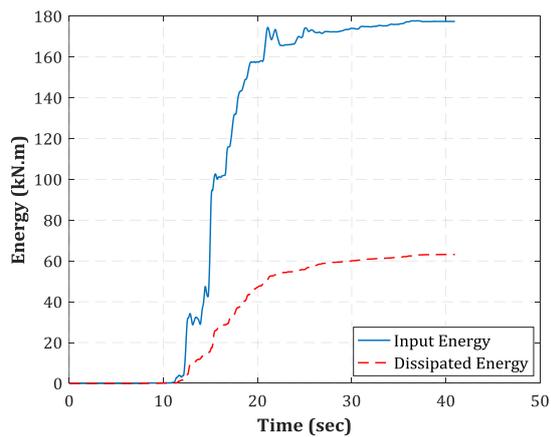
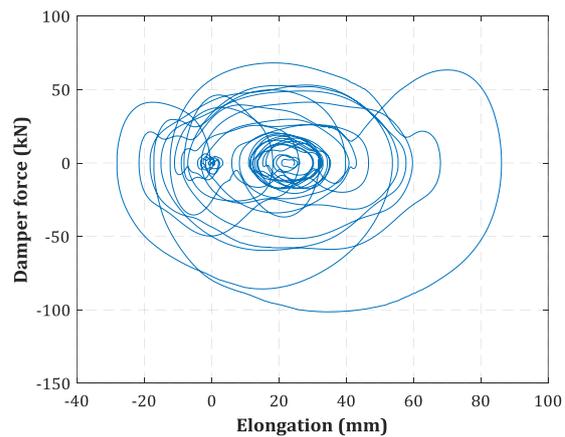

**(b)**

**Fig. 12:** Total energy dissipated by VDs in studied structure, and hysteretic response of VD in the 2nd story under KOBE/SHI000 record: (a) BSE-1 and (b) BSE-2 hazard levels.

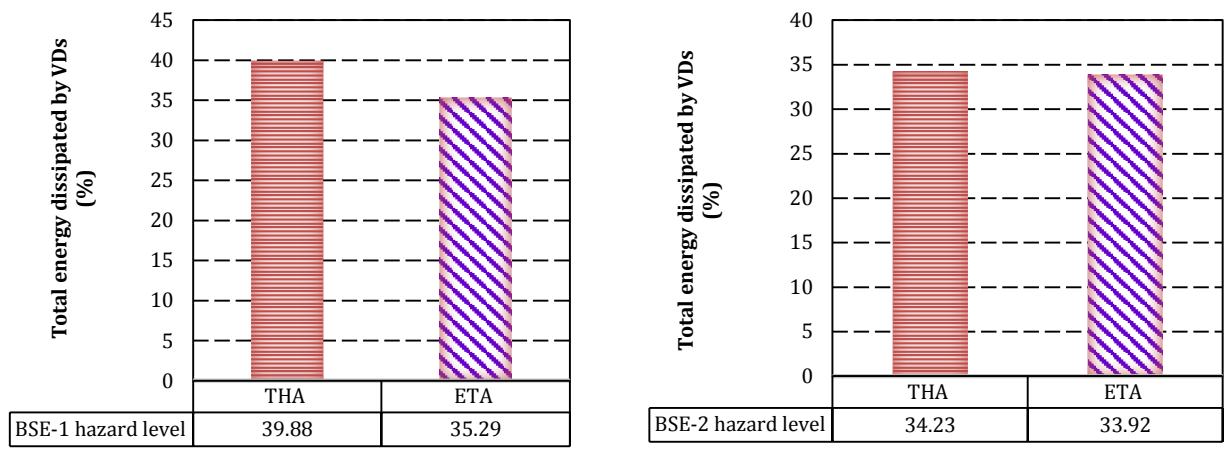

**Fig. 13:** Total energy dissipated by VDs (%) in studied structure based on THA and ETA at different hazard levels.

## 7. Summary and conclusions

In the ET method, structures are subjected to predesigned intensifying excitation function and their performance is assessed based on their response at different excitation levels. In this method, a single THA presents the structural performance for a continuous IMs range.

In this paper, the ET method is used to ascertain the optimal distribution of viscous dampers in a weak structure to reach the desired performance at different hazard levels, simultaneously. The viscous damper is one of the energy dissipation devices which can dissipate a large amount of seismic input energy to the structure. To this end, hysteretic energy compatible ETEFs are utilized and the validity of the results is examined by comparing them with the results achieved from a suite of ground motions. To optimize the placement of the dampers, the genetic algorithm is employed. The damping coefficients of the dampers are regarded as design variables in the optimization procedure and determined such that the sum of them has the least value. The distribution of dampers along the height of the structure is allocated utilizing the genetic algorithm in such a way that the structure satisfies allowable limits of the code in two performance levels of LS, and CP with the minimum sum of requisite damping coefficients of the dampers. Considering ETA20kd01-05 excitation functions, the following results are listed:

- The ET estimation of maximum IDRs compared to another set of the horizontal components of the earthquake records is also acceptable. Consequently, the sensitivity of maximum IDRs to another set of ground motions is comparatively low.
- It is concluded that after rehabilitation by optimally distributed dampers, the performance of the structure improves.
- For validating the results of the ET method, the difference between the maximum IDRs obtained by ETA and THA is calculated. This difference is greater in higher seismic intensities while the difference between the energy values is greater in lower seismic intensities.
- A more precise estimate of the differences between the ETA and THA results needs further research by increasing the number of stories and applying different sets of earthquake records.
- To have a better comparison between the energy values achieved by ETA and THA, the percentage of total energy dissipated by dampers for hazard levels of BSE-1 and BSE-2 is calculated. These differences for the ETA results compared to THA results are equal to (-11.52%) and (-0.91%) for hazard levels of BSE-1 and BSE-2, respectively. Thereby, the differences seem more reasonable.